\pdfoutput=1
\documentclass[10pt, pre,twocolumn,aps,showpacs,floatfix,nofootinbib,
superscriptaddress,longbibliography]{revtex4-2}
\usepackage{subfigure}
\usepackage{amssymb}
\usepackage{amsfonts}
\usepackage{amsmath}
\usepackage{amsthm}
\usepackage{epsfig}
\usepackage{graphicx}

\usepackage[usenames,dvipsnames]{color}
\usepackage[latin1]{inputenc}

\usepackage[left]{lineno}
\usepackage[hidelinks,unicode=true]{hyperref}
\hypersetup{colorlinks=true,
 	linkcolor=blue,
 	urlcolor=blue,
 	citecolor=blue,
 	pdfhighlight=/N
 }

 \usepackage{comment}
 \usepackage[normalem]{ulem}

\begin{document}

\title{Field emission tunnelling as a window onto fundamental issues in quantum mechanics}

\author{Richard G. Forbes}
\email{Permanent alias for author to whom correspondence should be addressed: r.forbes@trinity.cantab.net; current institutional e-mail: r.forbes@surrey.ac.uk .}
\affiliation{Quantum Sciences Group, School of Mathematics and Physics, University of Surrey, Guildford, Surrey GU2 7XH, UK.}

\begin{abstract}

\textcolor{blue}{\textit{Submitted as Chapter 19, in: T.C. Tuck (Ed.), Quantum Mechanics: A Century Later: In Celebration of Heisenberg's Seminal 1925 Paper. (To be published by World Scientific, Singapore, in early 2026.) v4, REVISED SUBMISSION AS ACCEPTED, with change of book title.}}
\bigskip

The physical processes of field electron emission (FE) and electrostatic field ionization (ESFI) both involve quantum-mechanical (QM) tunnelling. They are two of the (perhaps) seven historical paradigm examples of QM tunnelling. A modern relevance of FE and ESFI is that each forms part of the theory of various modern technologies and scientific techniques. However, there are fundamental quantum aspects of both FE theory and ESFI theory that are not fully understood.  This chapter aims to identify some of these fundamental difficulties and to offer some ideas for wider discussion.
\bigskip

\end{abstract}
%\pacs{85.45.Db, 85.45.Bz, 85.85.+j}

\maketitle

\tableofcontents

\section{Introduction}
\label{1}
\subsection{Background}
\label{1.1}
This chapter relates to the theory and problems of two field-induced processes that can take place near the surface of a metal or other material, particularly when the material has the form of a sharp, needle-like solid ``emitter" with a rounded apex, typically of radius of order 100 nm or less. (This shape facilitates electrostatic field enhancement at and around the emitter apex.)

Field electron emission (FE) occurs when an electron is extracted from a negatively charged emitter by the presence of a strong local surface electrostatic (ES) field, of magnitude typically of order a few V/nm, directed towards the surface. This field can be described as a ``negative conventional (textbook) surface ES field", and can be denoted by the negative parameter $E$. The extraction mechanism is electron tunnelling. The simplest theoretical case (of interest here) is tunnelling from a metal surface into vacuum, with the metal modelled by a Sommerfeld-type free-electron model.

By longstanding convention, the field-like parameters appearing in FE-related equations, such as eq.(6) below, are made positive. Hence, for use in equations, it is convenient to define a positive parameter $F$ as equal to $-E$.

Electrostatic field ionization (ESFI) is the process whereby a gas atom or molecule is ionized by a very strong local surface ES field close to a positively charged ``emitter". In this case the field is directed away from the emitting surface, and can be described as a ``positive conventional surface ES field". Its value is typically in the range 20 V/nm to 60 V/nm for the noble gases (but depending on the ionization energy of the species involved). 

When the gas atom or molecule is in space well above the emitter, the electron tunnels into free space, towards the emitter, and ends up in the emitter. The resulting positive ion is then repelled from the emitter surface.  When the gas atom or molecule is sufficiently close to the emitter then the electron can tunnel directly into a metal electron state above the emitter Fermi level. (The exclusion principle inhibits tunnelling into states below the Fermi level.) In scientific instruments based on ESFI, such as the field ion microscope, the dynamics of the operating gas prior to ESFI can be of importance, but this is of limited relevance here.

A distinction needs to be drawn between ESFI and so-called ``strong field ionization", where the ionization is induced by the oscillating electric component of an electromagnetic wave. Although related, these processes are best treated as physically different, with our interest here being primarily in ESFI.

\subsection{Motivation}
\label{1.2}
FE and ESFI are of scientific and technological significance for three main reasons.

(a) 	They are two of the (perhaps) seven historical paradigm examples of quantum tunnelling.

(b) 	They are part of the theory of various modern technologies, both implemented and proposed (speculatively in some cases), and of many of the ``machines of nanotechnology". A relevant list is given in Appendix A.

(c)	FE may be part of the cause of electrical breakdown in vacuum or in air or in other gases. For example, electrical breakdown might cause failure of the X-ray sources in medical scanners, or limit the performance of high-voltage particle accelerators, or be part of the physics of lightning strikes or lightning-strike protection. The detailed physical mechanisms of electrical breakdown are not yet fully understood in all cases.

There are also two specific reasons for future interest in FE theory. The field electron (emission) microscope can now image individual electron bonds at the apex of a carbon nanotube \cite{C1} (see Fig.1). Thus, opportunities for careful fine-scale sub-atomic experiments and comparisons with theory may exist in due course (though no experiments of this type have yet been carried out). Second, it is now established \cite{C2} that DNA mutation is facilitated by the quantum tunnelling of a hydrogen-related entity. Any future improvements in FE tunnelling theory, as discussed below, might conceivably have wider implications in due course, including in quantum biology and in the future subject of quantum botany.
\begin{figure}[!ht]
\includegraphics [scale=0.24] {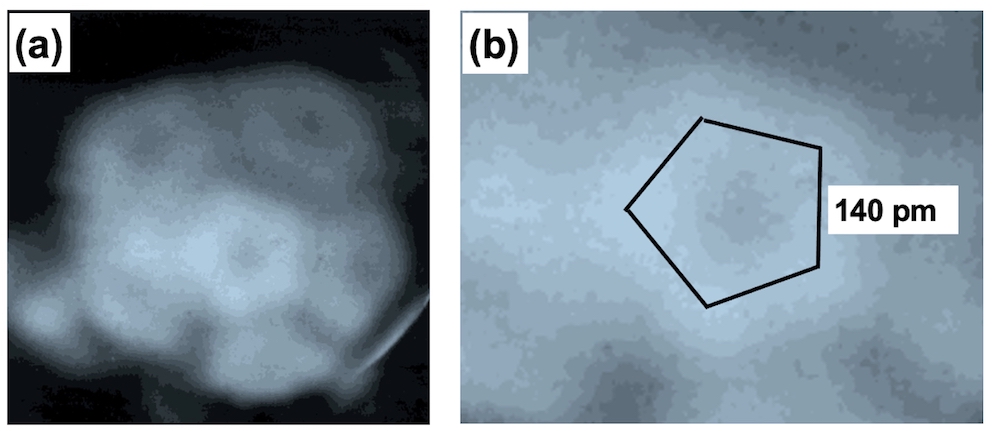}
\caption{(a) Field electron micrograph showing electron bonds associated with a five-membered carbon ring at the apex of a carbon nanotube. (b) Enlarged version of the central ring, with 
carbon nucleus separation marked. (Adapted from Fig. 2(a) in Ref. \cite{C1}.) Micrographs were taken at room temperature. Approximate magnification is of order $10^8$. Copyright (2000) The Japan Society of Applied Physics.}  
\label{Fig1}
\end{figure}

Quantum-mechanical theories of FE and ESFI have been under development for around 100 years but are still far from fully in place. The late President of the UK Institute of Physics (Marshall Stoneham) considered that developing an accurate theory of field electron emission was amongst the most difficult unsolved problems of theoretical physics \cite{C3}. The present author is not in a position to agree or disagree with this view, but has worked in FE and ESFI theory for over 50 years and considers that deep fundamental problems do exist in both subjects, and that these problems may link to wider issues in discussing and applying quantum mechanics.

This chapter aims to describe some of these perceived problems, and to offer some preliminary suggestions about how to discuss them. The aim is to initiate a wider discussion about specific points of interest in the context of FE and ESFI theory and the related technologies. E-mail responses from readers with more-specialized quantum-mechanical knowledge than the author has would be welcome.

Part of the thinking behind this chapter is as follows. If a quantum-mechanical problem has remained incompletely solved for nearly 100 years, then perhaps it is timely to discuss ideas that may be unorthodox from the point of view of conventional quantum mechanics. Another part of thinking is that, before developing the detailed mathematical models that will eventually be needed, it will be useful to first discuss qualitative issues and then, where possible, discuss simple mathematical models, even if these are necessarily inaccurate.

\section{Alternative utilisations of quantum mathematics}
\label{S2}

For convenience, the term ``quantum mathematics" is used here to describe the formalisms and related mathematics of quantum mechanics. Also, to make discussion more definite, focus is on describing the behaviour of electrons and (in order to avoid mathematical and conceptual complications) on considering only the simplest situations that involve the behaviour of an electron in space.

The author's perception is that, in the high-field-magnitude physics of interest here, quantum mathematics is applied to electron behaviour in two distinctively different ways. In what might be called the ``matter-distribution utilisation", quantum mathematics is used to describe the distribution in space of electron matter (or, in some special cases, such as electrons travelling in free space, the time-dependence of this matter distribution). For example, if we wish to calculate the ES field at some point in space close to a strongly polarized atom, then the atom's electron-wave-function-structure provides necessary information about the ``source" dipole-like charge distribution to which classical electrostatics is applied. At a more basic level, simple cartoons of atomic orbitals are widely used in teaching quantum mechanics and in qualitative discussions in chemistry and materials science.

It can be argued that in this utilisation there is a sense in which the electron wave-function describes the ``real" behaviour of the electron in the context of interest. This utilisation is close to Schr{\"o}dinger's original interpretation of the wave-function (see eq. (18) in \cite{C4}), subsequently abandoned in favour of what became called the ``Copenhagen interpretation". (Also see further comments below.)

Alternatively, what might be termed the ``pathway-choice utilisation" is used in situations where there is a choice of paths that an electron in a ``non-stationary" state might take. In this case, quantum mathematics is used to evaluate the probabilities that different paths will be followed. It is presumed that, in some contexts, different pathways to the same final state may be coherent: in such cases probability amplitudes (rather than probabilities) would need to be summed.
\begin{figure}[!ht]
\includegraphics [scale=0.19] {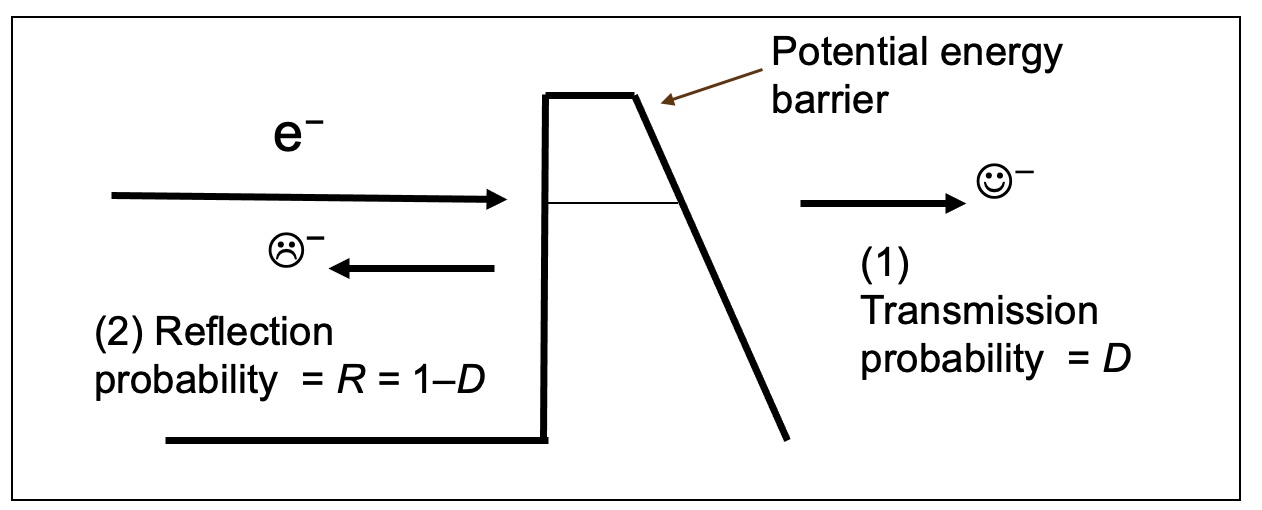}
\caption{To illustrate schematically that an electron approaching a potential-energy barrier has a choice of two pathways: (1) transmission; or (2) reflection.} 
\label{Fig2}
\end{figure}

A specific example of the pathway-choice utilisation is the tunnelling process in field electron emission, illustrated schematically in Fig. 2. Here, one first creates a travelling wave-function that represents an electron approaching the barrier. Applying quantum mathematics generates a situation where the total output of the mathematics includes both a travelling wave-function representing an electron transmitted through the barrier and a travelling wave-function representing an electron reflected from the barrier. The barrier transmission probability is found by comparing the squared moduli associated with the incoming and transmitted electron wave-functions.

Since it must presumably be a fundamental principle of physics that ``real electrons" cannot split into parts, it follows that the ``pathway-choice utilisation" of quantum mathematics is not a full description of the behaviour of ``real electrons".

This thought leads on to an apparent fundamental difficulty with FE tunnelling theory. It is well established that the escaping electron certainly has an interaction with the emitter electrons that is in the nature of a correlation interaction (or probably an exchange-and correlation (E\&C) interaction). However, due to the extreme theoretical difficulty of doing otherwise, this interaction is usually modelled as the classical image potential-energy associated with a classical point electron situated at a point in space outside but close to the emitter surface. A justification for doing this can be given \cite{C5,C6}. In the context of the presumed wave-like (and hence ``distributed") behaviour of real electrons, this point in space can presumably be taken to represent the centre-of-charge associated with the wave-function of the escaping real electron. The outcome (when the Sommerfeld model is used to model a planar metal-like emitter) is that the tunnelling potential-energy (PE) barrier is modelled as the so-called Schottky-Nordheim (SN) barrier (see Appendix B). Quantum mathematical methods for evaluating transmission probabilities for the SN barrier, at least approximately, are well established (see Appendix B).

But what if one wants a more accurate calculation of transmission probability? Presumably we would need to know ``as exactly as possible" how the correlation (or exchange-and-correlation (E\&C)) contribution to the PE barrier varies with the position of an escaping electron's centre of charge, as the escaping electron moves through the barrier. If there is any established method of doing this reliably within the framework of existing quantum mechanics, then the author is not aware of it. It would seem that any suitable theory would, in some way, have to disregard the possibility of wave-reflection effects during escape. Thus, the pathway-choice utilisation of quantum mathematics would presumably not work at this detailed level.

It is also not obvious how one would formulate even a qualitative picture of what is ``really" happening to an escaping electron. The author has a (very) rough picture of a tunnelling electron as behaving something like an octopus squeezing through a crack. A travelling electron, extended in space in the direction of travel, but of finite length, comes up to the barrier. When the front of the electron encounters the barrier then the whole electron slows down, and ``amplitude" gradually transfers through the barrier. On reaching the further side of the barrier, the transmitted part of the electron is presumably subject to the applied electrostatic field, which presumably tends to pull it away from the barrier. On the face of it, there appears to be a question (not commonly asked) of ``precisely why doesn't the electron break apart in such circumstances -- what keeps the electron together?"

In this discussion of alternative utilisations, there is also a question of what is meant by a ``quantum-mechanical (QM) measurement". The author's short answer is that QM measurement is \textit{observed} pathway choice, in an experimental context established by a human experimentalist. The author's thinking is that the observation can be made either by a human observer or by a machine that ideally is able to record the time and results of its observation process.

\section{Wave-function interpretation and the linguistics of quantum mechanics}
\label{3}
\subsection{Alternative interpretation for a ``material wave-function"}

An issue in the history of quantum mechanics has been the ``interpretation of the wave-function". The author's view is that the different utilisations of quantum mathematics require different interpretations for the squared modulus of the wave function. I also have the view that clear understanding is often impeded by the language used to discuss these issues, which sometimes seems to unnecessarily involve the classical concept of a classical point entity (in the present discussion a classical point electron).

For simplicity, consider the case of a hydrogen atom. Let the related conventional wave-function be
$\psi(r)$, where $r$ is the electron-matter position vector, measured relative to the position of the atomic nucleus. A widely encountered interpretation of the wave-function (e.g., see comments on Born rule \cite{C7}) is that its squared modulus $\psi(r)\psi{^*}(r)$ yields the probability density for measuring the electron as being at position $r$. As discussed below, I have difficulties with this
form of interpretation.

In the context of the ``matter distribution" utilisation of quantum mathematics, the following alternative is proposed, for discussion purposes at this stage. In the context of the International System of Quantities (ISQ) \cite{c8}, let $n_1$ denote the amount-of-substance of a system that contains one specified entity, in this case an electron. The parameter $n_1$ is a fundamental atom-level constant that plays the same role with respect to the ISQ quantity amount-of-substance as the elementary charge $e$ does for charge and the unified atomic constant mass constant $m_u$ does for mass.

If we further introduce the name ``entity" as the (provisional) name of an atomic-level unit of amount-of-substance, one can write
\begin{equation}
{n_1 = 1 \ \rm{entity}}.
\label{E1}    
\end{equation}
This unit, the ``entity", plays the same role, for amount-of-substance, as the atomic-level unit the Dalton does for mass.

If we introduce the further rule that (when no confusion would occur) the word ``entity" can be replaced by the name of the entity of interest, then in the present case we have
\begin{equation}
{n_1 = 1 \ \rm{electron}} .
\label{E2}    
\end{equation}

Using $n_1$, one can define a new wave-like quantity $\it{\Psi}_n(r)$ and its complex conjugate
$\it{\Psi}_n{^*}(r)$ by the equations
\begin{equation}
{\it{\Psi}}_n(r)={n_1}^{1/2} {\psi}_n(r); \\\\\ {\it{\Psi}}_n{^*}(r)={n_1}^{1/2} {\psi}_n{^*}(r) .
\label{E3}    
\end{equation}
where ${\psi}_n{^*}(r)$ is the complex conjugate of ${\psi}_n(r)$. From which it follows that:
\begin{equation}
{{\it{\Psi}_n(r) \it{\Psi}_n{^*}(r)} = n_1 {\psi}(r) {\psi}{^*}(r) ;}
\label{E4}    
\end{equation}
\begin{equation}
{{\int\it{\Psi}_n(r) \it{\Psi}_n{^*}(r) {\rm{d}}V} = n_1 = 1 \ \rm{electron} . }
\label{E5}    
\end{equation}
where ${\rm{d}}V$ is the element of volume. In the context of these equations,
${\it{\Psi}_n(r) \it{\Psi}_n{^*}(r)}$ can be interpreted as the (amount-of-substance) concentration of electron matter at position $r$.

Similar procedures, but using $e$ or $m_e$ instead of $n_1$, would lead to squared moduli that can be interpreted as the electron charge-density magnitude or the electron mass density at position $r$. For example, the charge-density interpretation of the wave-function is widely used to illustrate the outcomes of modern density functional theory (DFT) calculations -- see \cite{C9} for a FE example.

Two advantages of the form of wave-function interpretation embodied in eq. (4) are as follows. 
First, it avoids the linguistic implication, implicit in the common statement of wave-function interpretation, that somehow the electron is a point-like entity capable of being found at a definite point in space. Second, it avoids use of the terms ``probability" or ``probability density", and the resulting awkward question ``probability of what?"

Obviously, probabilities are essential in the pathway-choice utilisation of quantum mathematics, but they seem a conceptual hindrance when discussing the matter-distribution utilisation of quantum mathematics. It would perhaps be possible to interpret the conventional squared modulus 
$\psi(r) \psi{^*}(r)$ as the ``probability density for finding electron matter at position $r$", but it seems a more precise statement to let $\it{\Psi}_n(r) \it{\Psi}_n{^*}(r)$ define the concentration of electron matter at position $r$.

More generally, one can interpret what is being proposed here as the extension, to the wave-function level, of the concept of ``amount-of-substance".  This was introduced (as a ``seventh SI base-quantity") as part of the 1970s reforms to the international system of measurement, and has subsequently been applied to ``count atoms" and to describe the rates of observable atomic-level processes (e.g., see: \cite{C10}, Section 3.2.4; or \cite{C11}).

\subsection{``Point electrons" and the presumed impossibility of negative kinetic energy}
\label{3.2}

As already noted, the commonly stated interpretation of  $\psi(r) \psi{^*}(r)$ tends to carry the implication that the electron ``really" is a point entity capable of being found at some point in space, as opposed to ``really" being a distributed entity that (in space-based formulations) can only be found-by-measurement in some distributed state defined by the measurement apparatus. For this commonly stated ``point-electron" interpretation, a particular difficulty arises when the point of interest is inside a tunnelling barrier. At such a point, a point electron would have negative kinetic energy. So this would raise the wider question of whether a real electron could have negative kinetic energy (ignoring issues related to the Dirac equation and positrons).

It is clear that a \textit{classical} point electron cannot have negative kinetic energy. But the issue of whether a real electron could have negative kinetic energy is, philosophically, a different question. My answer to this question is ``certainly not", but it is not completely obvious whether this is an axiom about the nature of the ``real world" or an experimental fact. (If it is an experimental fact, rather than an axiom, what precisely are the experiments that show this?)

More generally, if a distributed (in space) electron traversing a barrier is to have positive kinetic energy overall then the related matter-distribution wave-function must ``stick out of the barrier" at one or both ends. Thus, if it is true that a ``real electron" (or other entity) cannot have negative kinetic energy, it follows that -- in the models of reality commonly used -- a real electron (or other entity) must be a distributed entity.

There are models of reality in which it is possible that this conclusion could  not be drawn, for example Bohm's ``pilot wave" model, in which -- according to some commentators (e.g. Norsen \cite{C12}) -- a point-like entity can cross a tunnelling PE barrier. But it seems that the conclusion certainly can be drawn in the commonly used models.

Obviously, behind the present discussion there lies the wider issue of whether physics should concern itself only with the results of measurements (see Wikipedia \cite{C13} for a useful introductory summary and further references). Those who adopt this view presumably might regard my ``matter distribution" interpretation of quantum mathematics as philosophically invalid. However, the widespread use of cartoons depicting electron orbitals and also the fact that we can use field electron microscopy (FEM) to ``see electron bonds" suggests that this interpretation is in fact empirically useful, even if currently we do not fully understand the physics of the FEM imaging process (see Section 5). There could also be the philosophical question of whether Fig. 1 could be considered to provide a rough measurement of electron behaviour.

As already has been noted, diagrams representing electron charge density are very widely used in papers describing the results of modern density functional theory.

\section{More linguistic and related issues}
\label{4}
Some other linguistic and philosophical issues seem to arise when real electrons are assumed to be distributed in space. In particular, it seems a meaningful question to ask: ``How long is a real electron, in the context under discussion?"  In high-energy physics an electron might be tightly localised and thus appear ``nearly point-like". However, in solid-state physics the electron length presumably might be determined by its coherence length, and presumably might be 20 nm or more. In ESFI, if the picture is valid that a real electron tunnels out of the atom over a period of time, then this would seem to imply that in some contexts the length of an electron might be a function of time.

There also seem to be linguistic issues relating to the way that the Heisenberg Uncertainty Principle is sometimes formulated. For a discussion in one spatial dimension $z$, a common form of statement is that the uncertainty $\Delta z$ in measuring the electron position and the uncertainty $\Delta p$ in simultaneously measuring the electron momentum $p$ are related by ${\Delta z} {\Delta p} \ge \hbar $, where $\hbar$ is the reduced Planck constant. But, again, what does the term ``electron position" mean if/when a real electron is 20 nm long?

An alternative might be to reformulate the uncertainty principle as a principle that relates minimum electron (etc.) kinetic energy to electron (etc.) length. The basic idea, as applied in one dimension, would be to consider an electron (or other entity) that is effectively confined to a region of space of finite length $L$. The smaller this length, then the higher the minimum electron kinetic energy $E_{\rm{min}}$.

The simplest illustration is an electron in a one-dimensional Sommerfeld well with infinitely high walls. The lowest energy state (corresponding to a half-cycle of a sinusoidal wave) has circular wave-number $k = {\rm{\pi}} / L $, and energy 
$E_{\rm{min}} = h_{\rm{P}}^2 / 8 m_{\rm{e}} L^2 $ where $h_{\rm{P}}$ is Planck's constant and $m_{\rm{e}}$ the electron mass. Obviously, an approach of this kind avoids using the concept of the position of a point electron.

For a numerical example in the above Sommerfeld-well case, consider a one-dimensional well of length 10 nm. The kinetic energy of the lowest-lying electron state is found (to 2 sig. fig.) to be 3.8 meV.

 It may also be arguable that the ``particle" aspects of the principle of wave-particle duality could be formulated in a clearer way. For some people, ``wave-particle duality" (as applied to matter) seems to mean that matter is not infinitely divisible, but comes in the discrete chunks that we now call ``particles" (but might be better called ``material entities" or by the old-fashioned name ``corpuscles"). (The extensive use of the term ``particle" by Newton, in his discussion of classical mechanics \cite{C14a,C14b}, may tend to associate point-like behaviour with the term ``particle".) With this interpretation of ``wave-particle duality", so-called particles are always (all the time) discrete material entities and are always (all the time) wave-like distributed entities.
 
For others, ``wave-particle duality" (as applied to matter) seems to mean that sometimes it is necessary to describe entity behaviour by quantum mathematics, but that sometimes it is more convenient to describe entity behaviour by the classical mathematics of point objects. A particular example is the Rutherford alpha-particle scattering experiments. Rutherford's original paper \cite{C15} used classical mechanics to analyse his results. But modern discussions might prefer to use the quantum mathematics of scattering theory.
It is arguable that the usual statements of wave-particle duality could usefully be replaced or supplemented by a statement about ``fundamental physical principles of existence and behaviour" (see Appendix C).

\section{The conceptual problems of ``seeing electrons" in the field electron microscope}
\label{5}

As already indicated, it is now possible to use the field electron microscope to image the bonds in a five-atom carbon ring at the apex of a carbon nanotube. The image in Fig. 1 probably had a diameter of order around 1 cm on the fluorescent phosphor screen. The screen image is a highly magnified and (probably) slightly blurred representation of the across-surface variations in the local electron emission current density emitted from the outside of the tunnelling PE barrier that holds the electron into the bond.

It can probably be assumed (in accordance with conventional thinking) that: (a) the observed images, as recorded by some form of electronic ``photographic" detector/recorder, are the product of the arrival at the screen of many individual electrons, one after another; and that (b) each electron causes light to be emitted from a phosphor grain (or a group of atoms) in an area of the screen that is very very much smaller than the area of the screen image. Thus, over time, from the time when the emitter is turned on by application of a step-like increase in applied voltage, the observed image changes from being very ``spotty" to having smoothly varying optical-intensity variations in the plane of the screen.

There is an apparently unsolved fundamental question as to precisely how this happens, given that the electron behaviour inside the bond must presumably be wave-like.
For an individual electron, what is incident on the inside of the PE barrier is presumably an ``electron wave" that is laterally coherent across the whole area of the bond. Mathematically, under the influence of the ES-field configuration between the emitter and screen, this wave presumably spreads out to cover the whole area of the screen image. But it would seem that the effect of the impact of the electron wave is localised to a very small screen area. Alternative conceptual possibilities (each with their own significant conceptual difficulties) would seem to be as follows.

If lateral localisation occurs at the microscope screen, then there has be some mechanism whereby all the charge and kinetic energy associated with the impinging widespread wave-front very quickly ``drains laterally" into one particular small location at the phosphor screen. To the present author, this does not seem plausible.

Alternatively, one might hypothesise that some mechanism acting when the wave emerges from the barrier or during its travel from emitter to screen causes the wave to become ``laterally highly localised". But the author is aware of no obvious mechanism to achieve this.

A thought is that at present we treat the interaction between an electron and the electrostatic field classically, which also means the field acts continuously. Possibly what is needed is a deeper physical understanding of this interaction (or, maybe, of the interaction with the sources of the field), and of how to treat these things ``in a proper quantum-mechanical fashion" (whatever that is). Feynman's approach to electron-electron interactions (e.g., \cite{C16}, in which electrons interact by exchanging virtual photons, might be one starting point for further thought.

It might also be worthwhile to attempt an experimental demonstration that would confirm the ``conventional thinking" described above.

This problem is not fully understood and is unsolved.

\section{Field emission tunnelling and the arrow of time}
\label{S6}

There is continuing interest in the question of how the arrow of time emerges from a quantum mechanical discussion of the behaviour of the world. (For a brief introduction and further references, see Wikipedia \cite{C17}.) It is usually considered that quantum mathematics is time-reversible, both in the wave-mechanics and in the matrix-mechanics implementations. Thus, it is sometimes argued, it is difficult to immediately see how a unique arrow of time can emerge from quantum mathematics. For example, a recent paper \cite{C18} suggests that two opposing arrows of time can emerge from a detailed study of dynamics.

However, the author's perception is that the quantum mathematics used in the pathway-choice discussion of FE tunnelling is \textbf{not} time-reversible as regards its statistical implications: rather it obeys/gives rise to a different fundamental principle of physics, namely transmission reciprocity. (This is a special case of the more general principle of scattering reciprocity, and can also be seen as an implication of the principle of detailed balance.)

For simplicity, consider the Schottky-Nordheim barrier discussed in Appendix B. For a given set of barrier parameters let the transmission probability for an electron approaching the barrier from the left be $D_{\rm{L} \rightarrow \rm{R}}$, with $ D_{ \rm{L} \rightarrow \rm{R}} < 1 $.  Transmission reciprocity means that the transmission probability $D_{ \rm{R} \rightarrow \rm{L}}$ for an electron approaching the same barrier from the right is equal to $D_{ \rm{L} \rightarrow \rm{R}}$, with
 $D_{ \rm{R} \rightarrow \rm{L}} < 1 $. It is easy to see that this is necessarily so within the JWKB approach to evaluating transmission probability: the probability depends on the JWKB integral which is the same whether you evaluate it from left to right or right to left. Hence, if you reverse time (and hence velocity) on a transmitted electron moving from left to right, and make it approach the barrier from the right, then it is not 100\% certain that the electron will go back to the left of the barrier. Hence the original change in electron state, from the left of the barrier to the right of the barrier, is ``not always time-reversible".

Clearly, if you simply reverse time in the thinking about the Schr{\"o}dinger equation, you could formally re-combine two ``partial electrons" travelling in opposing directions into a single electron travelling to the left, but this formal mathematics does not (in my view) correspond to ``real" electron behaviour.

Even when an electron approaches the barrier with energy sufficient to cross the barrier above its peak energy level, significant wave-mechanical reflection can occur, and similar arguments to those above apply.

In principle, it looks as if this type of argument could be extended further, to discuss the time-dependent behaviour (including the occupancy of different system parts) of a large system containing many individual non-time-reversible pathway choices. It is to be presumed that the statistical behaviour of such a system would not be time-reversible.

In general terms, the mechanisms just discussed seem to constitute one possible way to cause the entropy of a system to increase with time, leading to the existence of an arrow of time. This is not assumed to be the only possible causal mechanism for the arrow of time, others being related to changes in coherence and/or entanglement \cite{C19,C20}.

\section{The conceptual problem of near-surface electrostatic field ionization}
\label{7}

Most field electron emission theory is currently based on Sommerfeld-type models that assume that emitters have classically smooth surfaces. An urgent problem of FE theory is to ``get atoms into the theory". The work of Oppenheimer in 1928 \cite{C21,C22} showed that, in principle, the transmission process in FE could be thought of as equivalent to ESFI of surface atoms. Thus, it is thought that a possible route into atomic-level FE theory is to look again at the theory of the ESFI of a hydrogenic ion in its ground state. (A hydrogenic ion has a single electron surrounding a nucleus with charge $Ze$, where $e$ is the elementary (positive) charge, as before, and $Z$ is the nuclear charge number. It is helpful to discuss the ESFI of a hydrogenic ion, rather than the ESFI of a hydrogen atom, because it is desirable to get the charge-number $Z$ into the mathematical treatment.)

ESFI theory is also relevant to the theory of gas field ion sources and of the field ion microscope, and to the process of field post-ionization (FPI) that occurs in the context of atom probe tomography (APT). In APT, FPI is used both to explain aspects of emitted ion dynamics and to estimate near-surface ES-field values.

A good starting point for any fresh look at the theory of ESFI would be an ISQ version of the treatment given by Landau and Lifschitz (LL) in their well-known textbook on quantum mechanics, first published in English in 1958 \cite{C23a}. The present author formulated (but did not publish) an ISQ treatment of this kind some years ago \cite{C24}; this treatment now needs updating, but the mathematical results are still thought valid. The third English edition \cite{C23b} of the LL textbook, published in 1977, is the most useful because it gives a Gaussian-system version of the formula for ESFI rate-constant, as well as the atomic-units proof given in the 1958 edition. This enables the LL Gaussian-system formula to be compared with ISQ-system formulae.

To discuss planar FE theory in a slightly improved way, Forbes and Deane have introduced a new special mathematical function (SMF) ${\rm{v}_{\rm{FD}}}(x)$, where $x$ represents the independent variable in the Gauss Hypergeometric Differential Equation and ${\rm{v}_{\rm{FD}}}(x)$ is a special solution of this equation (for example, see \cite{C25}).

This function ${\rm{v}_{\rm{FD}}}(x)$ can also be applied to line-integrals used in ESFI theory and it can be argued that the resulting mathematics is superior to that used by Landau and Lifschitz. LL note that their treatment is expected to be valid only in the limit of low electrostatic (ES) fields. For a hydrogen atom in free space, both treatments generate the same result in the limit of low ES fields, but the new treatment is expected to be slightly more accurate at higher fields.

Part of the reason seems to be as follows. The new SMF ${\rm{v}_{\rm{FD}}}(x)$ is one of small group of SMFs that require TWO infinite series in order to provide a series definition. For such functions the process of Taylor expansion does not work well. The LL treatment of ESFI involves Taylor expansion and thus might be expected to not work well in mathematical detail.

However, aside from purely mathematical issues, there seem to be fundamental physical issues. The combination of the applied linear ES field and the Coulombic field due to the ionic nucleus generates a set of curved ES field lines that can usefully be described in parabolic coordinates. Close to the ionic nucleus there is a region where a classical point electron would have positive kinetic energy and ``downfield" of the atom there is also a region where the point electron would have positive kinetic energy.

In between these regions there is a PE barrier region. The ES-field distribution in this region might be expected, in some sense, to guide a real electron through the barrier. The requirement for conservation of electron angular momentum about the ionic nucleus would in reality complicate accurate physical and mathematical discussion; for present purposes we can disregard this complication. Because of the large difference in mass between the electron and the ionic nucleus, we also disregard issues relating to the need to use reduced mass.

The central issue is how to set up the physics for calculating the electron transmission probability through the barrier region. LL use parabolic coordinates to define an infinite number of possible ``paths" through the barrier. LL then consider each path and (using a JWKB-type theory) calculate relatively how much contribution to the final result is made via this path. An integration over a downfield plane normal to the applied field, taking into account what the classical velocity (in this plane) of a point electron would be, yields the total probability of ESFI per unit time, i.e. the ESFI transmission rate-constant.

There could be questions about the detailed validity of this approach, but the procedure may work more-or-less adequately when the hydrogenic entity is in free space. However, there is an issue of procedure when the entity is close to a metal surface and there is an exchange-and-correlation interaction (modelled as an image interaction) between the tunnelling electron and the surface. The issue is where to place the image charge. If one has a conceptual theoretical model in which quantum mechanics represents the outcome of an ensemble of alternative electron paths, each of which is traversed by a point-like ``full electron", then the simple position for the image charge is opposite the instantaneous position of the nucleus.

On the other hand, LL seem to be implementing a model of reality in which an electron is divided into infinitesimally small elements, each of which traverses a different path through the barrier and has its own path-related transmission behaviour. If one assumes, for the purposes of generating an E{\&}C interaction, that these elements act coherently, the logical position for the image charge would seem to be opposite the ion nucleus. The question of where to put the image charge is currently unresolved and seems to deserve wider discussion.

\section{Tunnelling-integral versus overlap-integral formulations of QM transmission theory}
\label{8}

In the above discussion, quantum-mathematical aspects of transmission theory have, where needed (as in Appendix B), been discussed using a tunnelling-integral formulation of transmission theory. Physically, this approach derives ultimately from the 1928 work of Fowler and Nordheim (FN) \cite{C26} on field electron emission (but note the detailed theory in this FN paper is seriously flawed -- for example see \cite{C27}). Virtually all later theoretical work on FE, including the important 1956 re-formulation of FE theory by Murphy and Good \cite{C28} (in order to take E{\&}C effects into account), adopts some variant of this underlying approach.

Tunnelling-integral-type approaches have also been widely used in ESFI theory, in particular: (a) in LL's theoretical discussion noted above; (b) as one approach in the useful review of hydrogen atom ESFI by Yamabe et al. \cite{C29}; (c) in discussions of how field ion microscopes work (for example, by Gomer \cite{C30}); and (d) in discussions of FPI theory (for example, by Kingham \cite{C31} and by Lam and Needs \cite{C32}).

An alternative basic approach is the overlap-integral type of formulation that derives ultimately from the 1928 papers by Oppenheimer \cite{C21,C22}. In this approach, separate Hamiltonians and wave-functions are established to describe electron motion in initial and final states in regions of space separated by the barrier. The transmission probability then depends on the degree of overlap of the exponentially decaying wave-function tails inside the barrier.

In the field ion microscope, ESFI takes place near and outside the so-called critical surface which (for a helium atom) is the locus of He-nuclear positions such that the energy level of the topmost electron orbital in the helium atom aligns with the emitter Fermi level. When discussing the imaging of an emitter surface atom, in circumstances where operating-gas distribution effects are not significant, the tunnelling-integral formulation suggests that the electron transmission rate-constant (ETR) primarily depends on the magnitude of the ES field between the critical surface and the emitter surface, whereas the overlap-integral formulation suggests that this ETR primarily depends on the distance between the critical surface and the nucleus of the imaged emitter surface atom. Thus, there is an opportunity to investigate how well the alternative formulations can explain the experimental details of field ion images.

A convenient discussion situation (used historically) is the helium ion imaging of a relatively small tungsten (111) crystal facet, as illustrated in Fig. 3. Two image features require explanation. The first is the fact that atoms in this facet can be resolved, which requires that the ETR for a gas-atom-nucleus position in the critical surface that is ``over a substrate atom" be sufficiently greater than the ETR for a critical-surface position that is ``over a point midway between atoms". The second is that the corner atoms in the facet image more brightly than the edge atoms between them.
\begin{figure}[!ht]
\includegraphics [scale=0.15] {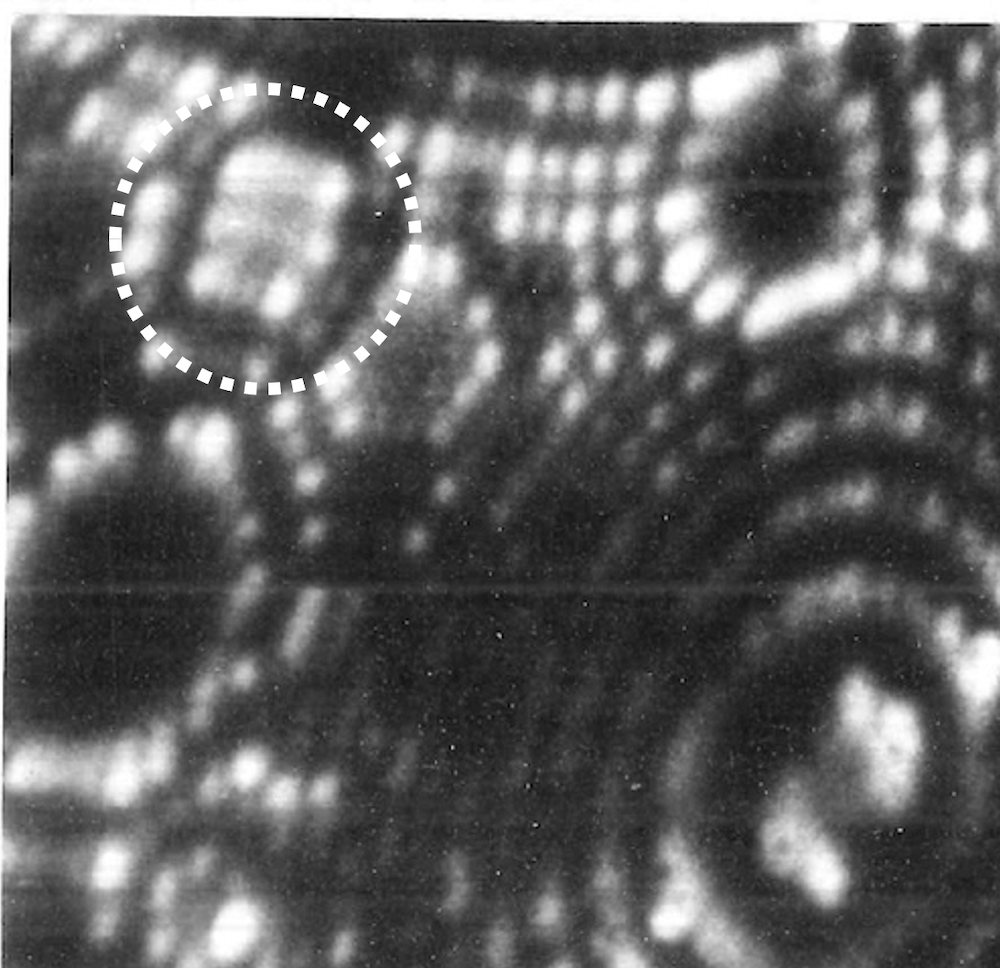}
\caption{Part of helium-ion image of tungsten emitter, taken near 80 K. A small W(111) crystal facet visible in the figure is circled. In the bottom atom-row, the corner atoms are imaged more brightly than the edge atom or the interior atom. (Adapted from Fig. A:31c in Vol. II of Ref. \cite{C33}.)} 
\label{Fig3}
\end{figure}

\begin{figure*}[!t]
\includegraphics [scale=0.20] {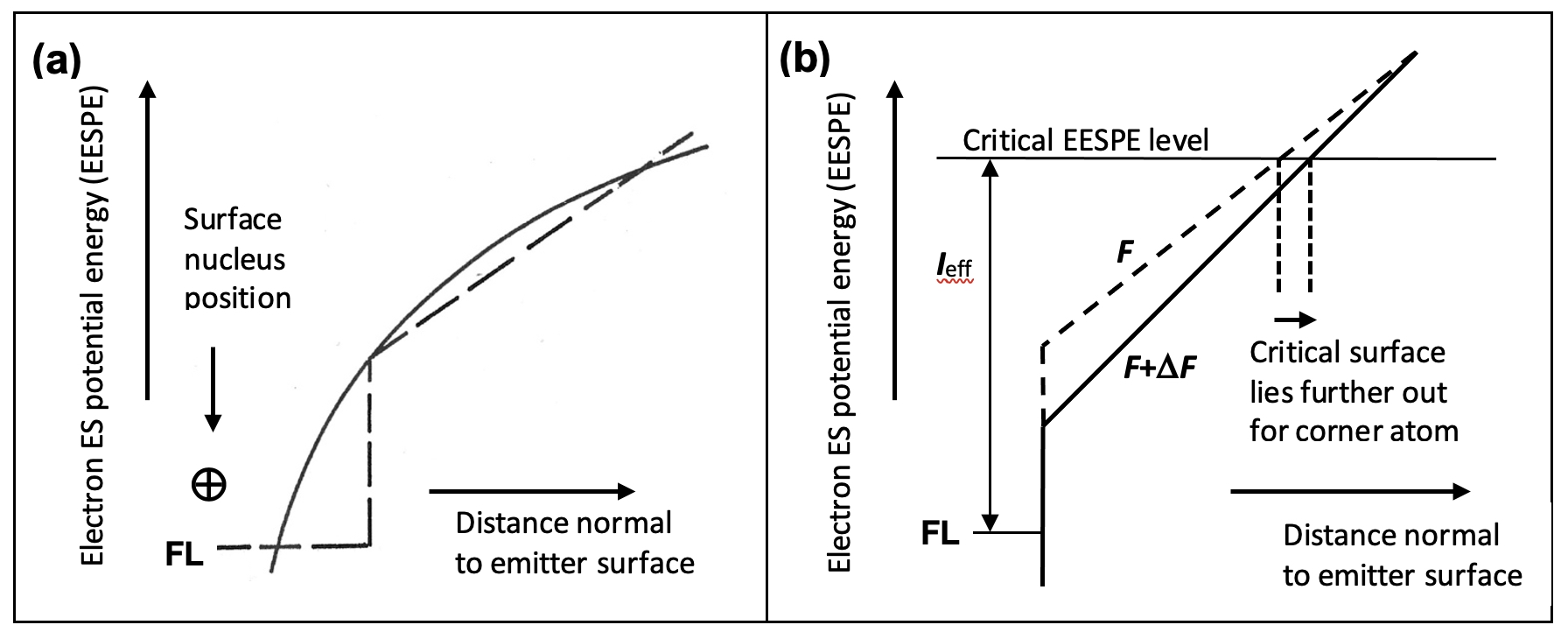}
\caption{Schematic diagram illustrating the modelling of electron electrostatic potential energy (EESPE) variations at a charged surface. (a) Fitting of a straight-line equivalent-barrier to the EESPE section through a surface-atom nucleus. (b) Straight-line equivalent barriers for an edge atom (broken line) and for a corner atom (full line). The extra positive charge (actually reduced electron charge) at the corner site means that the ES field is higher there and the critical surface is further out.} 
\label{Fig4}
\end{figure*}

By 1985, work by Homeier and Kingham \cite{C34} and by the present author \cite{C35} had eventually shown that, in the tunnelling-integral formulation, the ES-field variations across the emitter surface (between the critical surface and the emitter surface) were sufficient to explain atomic resolution.

Earlier, in the 1970s, it had been found \cite{C36,C37,C38} that overlap-integral formulations apparently could not provide a satisfactory explanation of field ion imaging. In the view of Goldenfeld et al. (\cite{C38}, p.338) this was because ``the behaviour of the (wave-)function in the complete potential(-energy) barrier...is essential in defining the value of the full probability flux through the barrier". An extensive theoretical re-investigation/review of the ESFI of the hydrogen atom in free space, carried out by Yamabe, Tachibana and Silverstone in 1977 \cite{C29}, reached a similar conclusion: namely that ``the barrier region dominates practical calculations, which are surprisingly sensitive to the accuracy of the wave-function there".

Returning to Fig. 3, the underlying physical reason why the corner atoms image more brightly is that (in accordance with Gauss' theorem) these atoms carry a higher effective positive charge than edge atoms or interior atoms (because there is less electron charge around a protruding emitter surface atom). In consequence, for the corner atoms: (a) the critical surface is further from the emitter nucleus than for edge and interior atoms; and (b) the field in the related barrier region is expected to be higher.

This effect is shown schematically in Fig. 4 below. Thus, qualitative expectation is that the tunnelling-integral formulation will be able to explain corner-atom brightness, but the overlap-integral formulation will not. Obviously. this conclusion is qualitative. It would be helpful to have it re-investigated using modern density-functional-theory techniques, but this has not yet been done.

Thus, the provisional conclusions from the above analysis of FIM image formation are that, in tunnelling contexts, tunnelling-integral formulations seem likely to be more reliable than overlap-integral formulations, and that (in some contexts) overlap-integral formulations may not even be able to predict results that ``go in the right direction".

Another significant difference between the two formulations discussed above is the following. In the formula for the electron transfer rate-constant (ETR), a parameter relating to the effective density of final electron states (in the emitter) currently appears in the overlap-integral formulation but not in the tunnelling-integral formulation.  Conceptually, if there are more states available for the electron to tunnel into than free-electron-type theory would predict, then the ETR is expected to be higher. And vice-versa.

This argument has recently been used \cite{C39} to make predictions about ``chemical contrast" in FIM images, i.e. the imaging of ``surface impurity" atoms in FIM images, albeit in the context of an overlap-integral formulation. (See this paper for one method of incorporating density-of-states effects into overlap-integral-type theory.) Thus, there is reason to think that effects of this general kind should probably be incorporated into tunnelling-integral theories. A brief summary of the essential merits of the two approaches is given in Table 1 below.
\begin{table*}
\caption{Comparison of different theoretical approaches to evaluation of
electron transfer rate-constants in the context of field ion imaging.}

\bigskip
\begin{tabular}{|c|c|c|}
\hline
\textit{Theoretical issue} & \textit{Tunnelling-integral theory} & \textit{Overlap-integral theory} \\
\hline
\hline
Can the theory qualitatively explain FIM images? & Yes & Apparently not \\ \hline
Does the theory currently contain a term & &  \\
relating to final-density-of-states effects & No & Yes \\
for non-free-electron metals? & & \\ \hline
\end{tabular}
\label{tab-models}
\end{table*}

This problem with tunnelling-integral formulations is more obvious in the situation where two solids with different non-free-electron-like band structures are separated by a penetrable asymmetric PE barrier that can be treated as planar. This is adequately the case with so-called ``Schottky-barrier devices", as discussed by the present author in a recent conference presentation \cite{C40}.

If the whole system of electrons (both sides of the barrier) is in joint thermodynamic equilibrium, then the principle of detailed balance must apply. In the case of states on either side of the barrier that have equal total-energy and equal normal-energy components, this seems to require the following: that (for these states) the expressions for the electron-transfer-rates from left to right and from right to left must \textit{both} contain density-of-states-related parameters corresponding to the different band-structures on \textit{both} sides of the barrier. The existing theory of Schottky-barrier devices appears not to meet this requirement \cite{C40}.

The conclusion drawn by the author from this Section's discussion is that neither tunnelling-integral formulations nor overlap-integral formulations are complete, correct and accurate in their present forms. It seems, as claimed by Stoneham \cite{C3}, that the problem of developing a fully correct theory of quantum tunnelling (and, more generally, of quantum transmission) is not yet fully solved.

\section{Conclusions}
\label{9}

This chapter has attempted a survey of what appear to be fundamental theoretical problems arising in the context of the theories of field electron and field ion emission, and has offered some suggestions about how to discuss them. The survey is not quite complete, since it does not cover fundamental problems relating to the process of field evaporation (FEV), in which surface atoms are removed from the emitter surface as ions, under the influence of a very high electrostatic field. The fundamental issue for FEV is precisely how the process of surface-atom ionization takes place at a highly charged surface. However, the author's view is that there are preliminary scientific issues (see \cite{C11} for a recent discussion) that need to be resolved before a discussion of more fundamental issues becomes timely.

The proposals that emerge from this survey can be classified in three groups, as follows.

\bigskip

\noindent \textit{Definite proposals}

\noindent (1) The issue of whether a ``real" material entity can have negative kinetic energy is an important preliminary fundamental question about models of reality and hence about quantum mechanics. The author's answer is ``certainly not", but there is a question of whether this answer is an axiom about the nature of physical reality or an experimental fact.

\noindent (2) When the answer to (1) is ``no", this implies that (specifically in the context of electron tunnelling, and presumably more generally) electrons -- and presumably all material entities -- are ``really" distributed entities rather than point entities.

\noindent (3) Given the above, it would probably be helpful if the language we use to discuss quantum mechanics (or the utilisation of quantum mathematics) were revised to remove all reference to ``the position of a (point) electron or other material entity". (Except where absolutely essential, as in Bohm-type theories.)

\noindent (4) As regards the tunnelling of material entities, neither of the main approaches (using tunnelling integrals or using overlap integrals) is currently complete, correct and accurate. There is an unfulfilled need to develop ``more accurate" theory (and there is almost certainly a need to develop more precise methods of comparing the outcomes of tunnelling experiments with theory). A ``more accurate" theory could likely be useful in various technological and industrial contexts. A ``highly accurate" theory of quantum tunnelling seems likely to eventually be needed as part of the theory of DNA mutation.

\bigskip

\noindent \textit{Other suggestions}

\noindent (5) It may be helpful to talk, not about ``alternative interpretations of quantum mechanics", but about ``alternative utilisations of quantum mathematics".

\noindent (6) It has been argued that, although the matter-distribution utilisation of quantum mathematics may sometimes be time-reversible, the pathway-choice utilisation is not. Hence, in a large system, the arrow of time seems not to be a mystery but could emerge from many individual pathway-choice events conceptually analogous to FE tunnelling. (But also in other ways, as well.)

\noindent (7) Where convenient, it might be clearer in principle to talk about the behaviour of ``material entities" (or the behaviour of photons) rather than about the behaviour of ``particles".

\noindent (8) There could be merit in replacing or supplementing the uncertainty principle by a principle that relates the minimum kinetic energy of a material entity to the length or volume of space to which it is restricted.

\noindent (9) There could be merit in replacing or supplementing the concept of wave-particle duality by a more comprehensive set of principles about the behaviour of matter and radiation, for example as suggested in Appendix C.

\bigskip

\noindent \textit{Unresolved fundamental issues}

\noindent (10) In the situation where a tunnelling electron has a correlation or exchange-and-correlation interaction with other electrons in the ``emitter", it appears unclear (even in principle) how to carry out an accurate calculation of the time-dependence of this interaction, and hence an accurate estimation of transmission probability.

\noindent (11) It is unclear how to explain precisely how the field electron microscope images the carbon bonds in a carbon nanotube.

\noindent (12) In the Landau and Lifschitz approach to the theory of electrostatic field ionization (ESFI), it is unclear how to deal correctly with image-charge effects when gas-atom ESFI takes place close to a conducting surface.

\noindent (13) There seems a fundamental question, not commonly asked, as to precisely why the front part of an electron emerging from a tunnelling barrier, and subject to the pull of the applied electrostatic field, does not ``pull apart" from the rest of the electron.

As noted in the introduction, the primary purpose of this chapter has been to offer suggestions for scientific discussion, with the specific aims of improving theories of field electron and field ion emission, but also in the hope they might be useful more generally. Responses, including informed criticisms, will be welcomed.

As regards procedure, for the most part all that can be done in the near future is to raise these suggestions in other forums. However, there are four proposals on which scientific work could reasonably begin as soon as time permits. I perceive the order of priority to be (most plausible first): (6), (9), (4) and (12). However, the issues I would most like to see solved are (10) and (11).

\bigskip

\section*{Appendix A:  Technologies potentially utilising field electron or field ion emission}
\label{CA1}

\noindent \textbf{\textit{Established technologies}}

\noindent *Electron microscopy, in its various forms

\noindent *Electron beam lithography

\noindent *Gas field ion sources \& related scanning ion machines

\noindent *Liquid metal ion sources

\noindent *Focused ion beam (FIB) machines

\noindent *Field electron microscopy \& related techniques

\noindent *Field ion microscopy \& related techniques

\noindent *Atom-probe tomography \& related techniques

\noindent *Field desorption mass spectrometry

\noindent *Electron sources for new ``fast" forms of medical X-ray \\ ``tomographic" machines

\noindent *Scanning probe lithography (in the field electron emission regime)

\medskip

\noindent \textbf{\textit{Technologies under development}}

\noindent *Electron sources for microwave generators (US military research)

\noindent *Many different emitting materials for large-area field electron sources

\noindent *Field desorption ion sources (which have been applied to fingerprint recognition)

\noindent *Field (ion) emission electric propulsion of spacecraft

\noindent *Space-vehicle neutralisers based on field electron emission

\noindent *New forms of imaging atom-probe

\noindent *Miniaturised mass spectrometers (e.g. for space applications)

\noindent *Miniaturised electron microscopes

\noindent *Ion implantation systems based on liquid metal ion sources

\noindent *Lightning-strike protectors

\noindent *Amorphous metal non-linear resistors (AMNR devices)

\medskip

\noindent \textbf{\textit{Technologies under exploration}}

\noindent *Nanoscale vacuum-channel and air-channel transistors

\noindent *The surface technique of ``Near Field Emission Scanning Electron Microscopy"

\noindent *Technologies based on microgas discharges

\noindent *Neutralisers for Hall thrusters for spacecraft propulsion

\noindent *New battery forms based on electrostatic energy storage (probably unlikely to happen)

\noindent *Silent drones driven by electrostatic propulsion

\bigskip

\section*{Appendix B:  The Schottky-Nordheim barrier}
\label{CAB}

The Schottky-Nordheim (SN) barrier is the potential-energy (PE) barrier above a smooth planar metal surface when the joint effects of an applied negative ES field of magnitude $F$ and classical image effects are taken into account. For a SN barrier with zero-field height equal to the local work function $\phi$, the difference $M(\phi,F,z)$ between the electron potential energy and the electron kinetic-energy component normal to the surface is
\begin{equation}
{M(\phi,F,z) = \phi - eFz - {e^2}/16 \mathrm{\pi} \epsilon_0 z}
\label{E6}    
\end{equation}
where $z$ is distance outwards from the metal surface, $e$ is the elementary (positive) charge and $\epsilon_0$ is the electric constant (aka the vacuum permittivity). In the simplest tunnelling-integral formalism (the so-called first-order JWKB formalism), the electron transmission probability $D$ for escape from the emitter is approximately given by
\begin{equation}
{D \approx \exp[-{(8 m_{\rm{e}} / {\hbar}^2)}^{1/2} \int_{z_1}^{z_2} M^{1/2}(\phi,F,z) {\rm{d}}z}
\label{E7}    
\end{equation}
where $m_{\rm{e}}$ is the electron mass in free space, $\hbar$ is the reduced Planck constant, and the integral is taken between the zeros $z_1,z_2$ of $M^{1/2}(\phi,F,z)$.

\bigskip

\section*{Appendix C:  	Suggested ``Fundamental physical principles of existence and behaviour"}
\label{CAC}

It is suggested that the principle of wave-particle duality could be replaced or supplemented by a set of
``fundamental physical principles of existence and behaviour". A tentative first draft for such a set is as follows. Emphasis is on content rather than precise wording, and ordering is provisional.

1)	Travelling electromagnetic (EM) radiation is not infinitely divisible: all EM radiation consists of travelling EM entities now called photons.

2)	Travelling photons are distributed in space and their motion is always wave-like.

2a)	A travelling photon has energy $h_{\rm{P}} \nu$, where $h_{\rm{P}}$ is Planck's constant and $\nu$ is the frequency of the EM radiation.

2b)	When a photon is absorbed by matter then normally the whole energy of the photon is transferred to the matter.

3)	Matter is not infinitely divisible: all forms of matter are composed of discrete entities that can be called
``material entities".

3a)	Material entities are always distributed in space and their behaviour is always wave-like.

3b)	However, in some circumstances it may be convenient to model the behaviour of a material entity as the motion of a mathematical point placed at its centre of mass or charge.

4)	In some circumstances, when a large number of material entities are interacting with each other, it may be convenient to represent the collective behaviour of some part of the assembly of entities as a ``smeared out" continuous distribution of matter or charge.

4a)	Examples of this are classical space-charge models and the use of ``jellium" in surface-physics models.

Due to the possible ambiguity in the interpretation of the word ``particle", its use is avoided in the above statement of principles.

A difference from the usual statement of wave-particle duality is the emphasis that photons and material entities are always (\textit{all the time}) discrete entities and are always (\textit{all the time}) distributed in space, and that their behaviour is always (\textit{all the time}) wave-like.

\bibliography{Bibliography-Window}

\end{document}